\documentclass[twocolumn,english,floats,showpacs,prb]{revtex4}
\usepackage[T1]{fontenc}
\usepackage[latin9]{inputenc}
\usepackage{array}
\usepackage{amsmath}
\usepackage{graphicx}
\usepackage{amssymb}

\makeatletter

\providecommand{\tabularnewline}{\\}

\@ifundefined{textcolor}{}
{%
 \definecolor{BLACK}{gray}{0}
 \definecolor{WHITE}{gray}{1}
 \definecolor{RED}{rgb}{1,0,0}
 \definecolor{GREEN}{rgb}{0,1,0}
 \definecolor{BLUE}{rgb}{0,0,1}
 \definecolor{CYAN}{cmyk}{1,0,0,0}
 \definecolor{MAGENTA}{cmyk}{0,1,0,0}
 \definecolor{YELLOW}{cmyk}{0,0,1,0}
 }

\makeatother

\usepackage{babel}

\makeatother

\usepackage{babel}

\begin{document}

\title{A real-space effective c-axis lattice constant theory of superconductivity}

\author{X. Q. Huang$^{1,2}$}

\email{xqhuang@netra.nju.edu.cn}

\affiliation{$^{1}$Department of Telecommunications Engineering ICE, PLAUST,
Nanjing 210016, China \\
 $^{2}$Department of Physics and National Laboratory of Solid
State Microstructure, Nanjing University, Nanjing 210093, China }

\date{\today}
\begin{abstract}
Based on the recent developed real-space picture of superconductivity,
we study the stability of the superconducting vortex lattices in layered
superconductors. It is shown that the effective c-axis lattice constant
play a significant role in promoting the superconducting transition
temperature in these materials. An unified expression $T_{c}^{\max}=10c^{*}-28$
can be applied to estimate the highest possible $T_{c}^{\max}$ for
a given layered superconductor with an effective c-axis lattice constant
$c^{*}$. For the newly discovered iron-based superconductors, our
results suggest that their $T_{c}$ cannot exceed 60 K, 50 K and 40
K for the 1111, 21311 and 122 series, respectively. In the case of
copper-based oxide superconductors, it seems that the highest $T_{c}$
can reach about 161 K without applying of the external pressure. In
our theoretical framework, we could interpret the experimental results
of the completely different superconducting transition temperatures
obtained in two very similar cuprate superconductors (La$_{2-x}$Ba$_{x}$CuO$_{4}$ of
40 K and Sr$_{2-x}$Ba$_{x}$CuO$_{3+\delta}$ of 98 K). In addition, 
the physical reason why the superconductivity
does not occur in noble metals (like gold, silver and copper) is discussed.
Finally, we argue that the metallic hydrogen cannot exhibit superconductivity
at room temperature, it even cannot be a superconductor at any low
temperature. 
\end{abstract}


\maketitle

\section{Introduction}

Finding the room temperature superconductors has been an elusive dream
of scientists in the field of condensed matter physics. In 1986, the
discovery of high temperature cuprate superconductivity by Bednorz
and M\"{u}ler \cite{bednorz} shocked the superconductivity community
and rekindled the dream of room-temperature superconductivity. Since
then, great efforts have been devoted to finding out new materials
with a higher critical temperature.\cite{mkwu} Many cuprate superconductors
have since been discovered and the highest temperature superconductor
was HgBa$_{2}$Ca$_{2}$Cu$_{3}$O$_{8+\delta}$ with $T_{c}=138$
K. Its $T_{c}$ can be increased as high as $164$ K under a high
pressure of about $30$ GPa.\cite{gao} Later, several materials have
been claimed to be room-temperature superconductors which have been
quickly proved to be false. However, how to enhance the superconducting
transition temperature is always one of central concerns in the field
of superconductivity.

In 2008, Japanese researchers discovered the superconductivity in
the iron-oxypnictide family of compounds.\cite{kamihara} Like the
Cu-based superconductors,\cite{bednorz,mkwu,gao} the new superconductors
have layered atomic structures and can conduct electricity without
resistance at relatively high temperatures than the conventional low-temperature
superconductors. Similarly, the new discovery has triggered intensive
research worldwide and the maximum critical temperature has been raised
from the initial report of $T_{c}=26$ K in LaO$_{1-x}$F$_{x}$FeAs
\cite{kamihara} to about 55 K in SmO$_{1-x}$F$_{x}$FeAs\cite{ren}.
Facing the rapid increase of $T_{c}$, some researchers seem too optimistic
about the new discovery and they even claimed that these compounds
could be useful for developing room-temperature superconductivity.
In fact, more than one year ago, we proved theoretically that the
maximum $T_{c}$ of the new iron arsenide superconductors is very
difficult to reach 60 K by means of elements substitution or by applying
hydrostatic pressure, and this prediction is still correct up to now.
As pointed out by Chu recently, \textquotedblleft The discovery of
iron-based pnictide superconductors may have reinvigorated the field
of high-temperature superconductivity, but the cuprate superconductors
are still in the game\textquotedblright.\cite{chu}

In addition, physicists considered that a high pressure can make the
hydrogen molecules into atoms and make it a metallic conductor. Moreover,
it was predicted according to the BCS theory\cite{bcs} in 1968 by
Ashcroft\cite{ashcroft} that solid metallic hydrogen may be superconducting
at high temperature, perhaps even room temperature. However, to get
metallic hydrogen is still a distant dream, not to mention the realization
of the superconductivity. Based on our theory,\cite{huang1,huang2}
we would like to point out that even if there exists metallic hydrogen,
it is also impossible to exhibit the superconductivity at any low
temperature.

In this paper, the discussion of enhancement of superconductivity
is presented in the framework of the real-space superconducting vortex
lattices. We show that the effective c-axis lattice constant play
a key role in promoting the superconducting transition temperature
in the layered superconductors.

\section{A comparison between two superconducting theories established in
momentum space and real space}

\subsection{BCS theory of superconductivity}

\begin{figure}
\begin{centering}
\resizebox{1\columnwidth}{!}{ \includegraphics{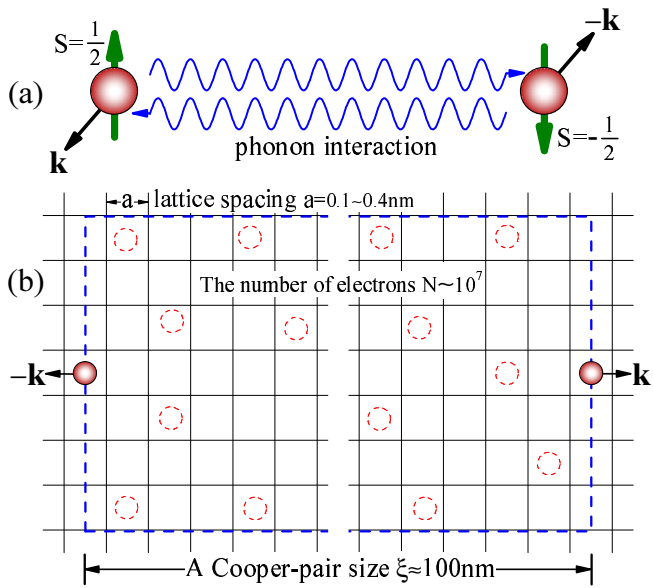}}
\par\end{centering}

\caption{(a) Two electrons are bound into a Cooper pair in the momentum space,
the paired electrons have opposite spin and opposite momentum and
gives rise to the superconductivity. (b) In the real space, the paired
electrons (one moving left, the other moving right.) can be separated
by a coherence length (the Cooper-pair size) of up to 100 nm. And
there may exist $10^{7}$ electrons (or electron pairs) between the
paired electrons.}

\label{fig1}
\end{figure}

In the framework of the BCS theory,\cite{bcs} the superconducting
current is carried by the so-called Cooper pairs of electrons which
are held together by lattice vibrations (an exchange of phonon) in
the material. The usual analysis in the BCS theory relies on a momentum-space
picture, the Cooper pairs are formed in momentum space ($\mathbf{k}$-space)
and the paired electrons have opposite spin and opposite momentum
($\mathbf{k}$$\uparrow$,$\mathbf{-k}$$\downarrow$), as shown in
Fig. \ref{fig1}(a). Of course, if the two electrons are to remain
in the same paired state forever, then they must undergo a continuous
exchange of virtual phonon in the BCS picture.

Although the physics community accepts that the BCS theory can successfully
describe the behaviors of most metallic superconductors, it is commonly
believed that the phonon based BCS theory is invalid for the description
of the behavior of the high-temperature superconductors. As pointed
out by Anderson,\cite{anderson0} the need for a bosonic glue (phonon)
in cuprate superconductors is folklore rather than the result of scientific
logic. It is well known that the superconducting electrons (as defined
by BCS theory) are a momentum space order phase, while at the same
time a disorder phase in real-space. From both the physics and mathematics
point of view, the fundamental differences between two descriptions
(one order and one disorder) of superconducting electrons imply that
the BCS theory may not be scientifically self-consistent. From the
perspective of the real-space correlations, two bound electrons (a
Cooper pair) can separate by a coherence length of $\xi$ up to 100
nm, which is also called the coherence length. It should be noted
that there are about $10^{7}$ randomly distributed electrons (or
pairs) between them in the conventional superconductors, as illustrated
in Fig. \ref{fig1}(b). More importantly, the Coulomb interaction
is the elementary electrical force that causes two negative electrons
to repel each other, hence, two important questions arise: (1) How
can the real-space repulsions among electrons (or pair-pair repulsions)
and the attractions among electrons and ions be eliminated to support
the formation of the Cooper pairs? (2) Normally, there are two kinds
of interactions in the superconducting materials, one is the very
strong short-range electron-electron interactions [see Fig. \ref{fig1}(b)],
the other is the rather weak long-range electron-phonon interactions
[see Fig. \ref{fig1}(a)]. Now, the question is why can the stronger
short-range electron-electron interactions be completely ignored in
the BCS theory?

\subsection{Wigner crystal and the real-space superconducting theory}

As we know there are many theories about the cause of the superconductivity
in cuprate high temperature superconductors. Unfortunately, most of
these theories contradict each other and they may be on the wrong
track as emphasized by Anderson\cite{anderson0}.
 In our opinion, the natural strong repulsion
between two electrons is impossible to be totally overpowered by a
lattice vibration (known as a phonon) and all superconductors should
share exactly the same physical reason. The present situation of one-material
one-mechanism of superconductivity should be changed. In other words,
a new and unified theory which can explain all superconducting phenomena
has proved more important and urgent today.

With increasingly better samples and advances in experimental techniques
(for example the Scanning Tunneling Microscopy\cite{hanaguri}), a
vast amount of data from these experimental studies reveal that the
superconducting electrons are more likely to congregate in some real
space quasi-one-dimensional rivers of charge separated by insulating
domains.\cite{hanaguri,kivelson,tranquada,komiya} Obviously, a correct
and reliable theory of superconductivity has to take into account
these new results. Although many researchers have been trying to replace
the conventional superconducting picture (dynamic screening) with
a real space picture, but how to construct a proper model related
to the formation of the one-dimensional charge rivers will still be
a major challenge for those devoting themselves to crack the mystery
of high-temperature superconductivity.

\begin{figure}[tp]
\begin{centering}
\resizebox{1\columnwidth}{!}{ \includegraphics{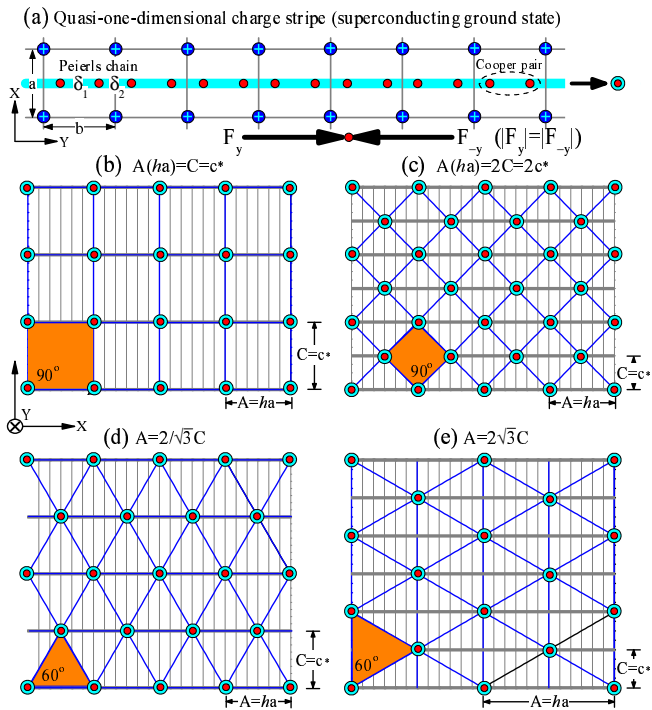}}
\par\end{centering}

\caption{(a) The short-range repulsive forces among electrons can be canceled
out by arranging them into one-dimensional real-space charge river
(charge stripe). A stable one-dimensional charge stripe is confined by the domain-walls
of the positive ions. It is proved that the electron-electron repulsions
inside the stripe can be eliminated completely as indicated in the
figure. (b)-(e) The long-range electron-electron repulsions among
different stripes can also be canceled out by arranging the stripes
into some periodic Wigner crystals. To achieve the highest superconducting
transition temperature, the superconducting vortex lattices should
be in the following four stable structures, (b) and (c) the vortex
lattices with tetragonal symmetry, while (d) and (e) having the trigonal
symmetry, where $h$ is a positive integer.}

\label{fig2}
\end{figure}

In recent years, we have tried to propose a real space superconducting
mechanism which may provide new insights into the nature of the superconductivity.\cite{huang1,huang2}
In our scenarios, all the superconducting electrons can be considered
as the `inertial electrons' moving along some quasi-one-dimensional
real-space ballistic channels, as shown in Fig. \ref{fig2}(a). In
the previous paper,\cite{huang2} it was proved theoretically that
a static one-dimensional charge stripe can be naturally formed inside
the superconducting plane and the Coulomb repulsion between electrons
can be suppressed completely, as indicated in Fig. \ref{fig2}(a).
According to the principle of minimum energy, for three-dimensional
bulk superconductors, the one-dimensional charge stripes can further
self-organize into some thermodynamically stable vortex lattices (the
Wigner crystal) with trigonal or tetragonal symmetry, as shown in
Figs. \ref{fig2}(c)-(f). It is not difficult to prove that the electron-electron repulsions among
different stripes can also be completely canceled out due to the symmetries in the Wigner crystals.

\section{the stability of the superconducting vortex lattices}

On the basis of our theory as described in Fig. \ref{fig2}, the superconducting
critical temperature of the layered superconductors is closely related
to the stability of the vortex lattices. Obviously, the stripe-stripe
interaction is the most important factor relevant to the stability
of superconducting vortex lattices. It is easy to find that the competitive
interactions between stripes can be greatly reduced by increasing
the stripe-stripe spacing and this, in turn, enhances the stability
of the superconducting state and increases the corresponding superconducting
transition temperature. For the layered superconducting materials,
there are usually two ways to control the stripe-stripe spacing
inside the superconductors. The first is the carrier concentration
and the second is the c-axis lattice constant of the samples. Normally,
a superconducting sample with a low carrier concentration and a large c-axis
lattice constant may have a higher superconducting transition temperature.
Of course, the too low carrier concentration is not conducive to the
formation of superconducting vortex lattices due to the lack of effective
competition among electrons.

\begin{figure}
\begin{centering}
\resizebox{1\columnwidth}{!}{ \includegraphics{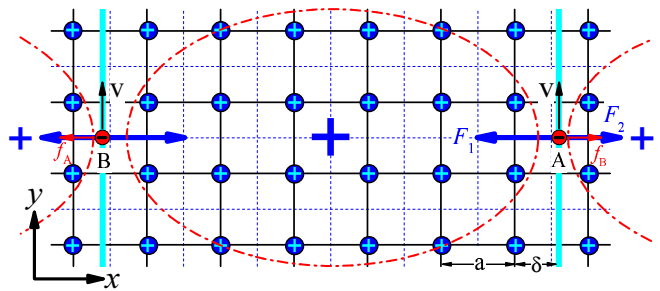}}
\par\end{centering}

\caption{When two electrons moving along the same direction ($y$-axis) with
the same velocity (or static, with the same $y$ coordinate ), the $x$-axis Coulomb repulsion between them can
be fully overcome ($f_{B}=F_{1}-F_{2}$) by the positive ion background.}

\label{fig3}
\end{figure}

To form a stable superconducting vortex lattice, both electron-electron
repulsive interactions within the charge stripes and between the stripes
should be inhibited thoroughly. The former situation was studied and
resolved in one of our previous papers,\cite{huang2} here we will
pay our attention to the later situation. It will be shown that the
nature of the electron-electron repulsive interaction can even be changed
into an attractive type because of the positive ion background.

Our theory is based on the experimental facts that the superconducting
electrons may be associated with a long-range spatial coherence charge
ordered state. In order to study the stability of the superconducting
vortex lattices of Fig. \ref{fig2}, we can consider the simplest
case where the stripe-stripe interaction can be simplified as the
electron-electron interaction, as shown in Fig. \ref{fig3}. We will
show that by adjusting the parameters $\delta$ the electron interaction
can not only be repulsive but can be also attractive. Moreover, there
is a special value of $\delta$ that can lead to a completely suppression
of the Coulomb repulsion, that is 
\begin{equation}
f_{B}=F_{1}-F_{2}\label{eq:force}\end{equation}

\begin{figure}
\begin{centering}
\resizebox{1\columnwidth}{!}{ \includegraphics{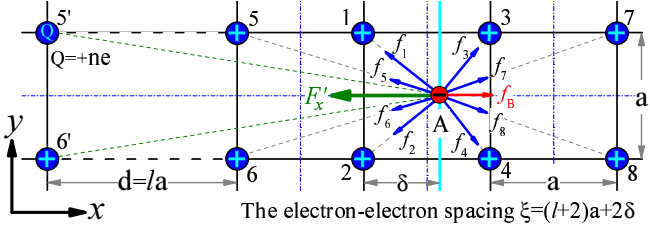}}
\par\end{centering}

\caption{A diagram illustrates how the Coulomb repulsion between two electrons
be eliminated by considering the electron-ion interactions. Note that
the electron $\mathbf{A}$ and $\mathbf{B}$ (which is not shown in
this figure) are located inside two adjacent stripes, respectively.}

\label{fig4}
\end{figure}

In the following, we will discuss how electron-electron repulsion
between the two adjacent charge stripes be overcome? It is worth noting
is that this is merely the well-known Coulomb screening effect of
the positive ion background. For simplicity, apart from the Coulomb
repulsion between two electrons, we consider only the nearest-neighbors
(both in $\mathit{x}$ and $\mathit{y}$-directions) and next-nearest
neighbors (only in $\mathit{x}$-direction) electron-ion interactions
in this study, as shown in Fig. \ref{fig4}. Here we analyze mainly
the forces applied on the electron $\mathbf{A}$ of Fig. \ref{fig4}
and the discussion is similar to another electron $\mathbf{B}$ (not shown in this figure).

According to Fig. \ref{fig3} and Fig. \ref{fig4}, the Coulomb forces
applied to the electron A can be divided into three parts. The first
part is electron-electron ($\mathbf{A}$ and $\mathbf{B}$) repulsion
which is given by

\begin{equation}
f_{B}=\frac{e^{2}}{4\pi\varepsilon_{0}\left[(l+2)a+2\delta\right]^{2}},\;(l=1,2,3,\cdots)\label{fb}\end{equation}
The second part is contributed by the eight ions (they are marked
by 1, 2, 3, 4, 5, 6, 7 and 8 in Fig. \ref{fig4}, and each ion carries
a positive charge $\mathit{ne}$) around the elect on $\mathbf{A}$,
they are \begin{eqnarray}
F_{x}^{(12)} & = & f_{1}+f_{2}\nonumber \\
 & = & -n\frac{e^{2}}{4\pi\varepsilon_{0}}\frac{\delta}{\left[(a/2)^{2}+\delta^{2}\right]^{3/2}},\label{f12}\end{eqnarray}
\begin{eqnarray}
F_{x}^{(34)} & = & f_{3}+f_{4}\nonumber \\
 & = & n\frac{e^{2}}{4\pi\varepsilon_{0}}\frac{a-\delta}{\left[(a/2)^{2}+(a-\delta)^{2}\right]^{3/2}},\label{f34}\end{eqnarray}
\begin{eqnarray}
F_{x}^{(56)} & = & f_{5}+f_{6}\nonumber \\
 & = & -n\frac{e^{2}}{4\pi\varepsilon_{0}}\frac{a+\delta}{\left[(a/2)^{2}+(a+\delta)^{2}\right]^{3/2}},\label{f56}\end{eqnarray}
\begin{eqnarray}
F_{x}^{(78)} & = & f_{7}+f_{8}\nonumber \\
 & = & n\frac{e^{2}}{4\pi\varepsilon_{0}}\frac{2a-\delta}{\left[(a/2)^{2}+(2a-\delta)^{2}\right]^{3/2}}.\label{f78}\end{eqnarray}
The resultant force of this part on the electron $\mathbf{A}$ can
be expressed as:\begin{equation}
F_{x}=F_{x}^{(12)}+F_{x}^{(34)}+F_{x}^{(56)}+F_{x}^{(78)}.\label{fx}\end{equation}
The third part is contributed by the other eight ions around the elect
on $\mathbf{B}$, the corresponding resultant force $F_{x}^{'}$ can
be written as:

\begin{equation}
F_{x}^{'}=-n\frac{e^{2}}{4\pi\varepsilon_{0}}\sum_{i=1}^{4}\frac{\delta+(i+2)a}{\left[(a/2)^{2}+\left[\delta+(i+2)a\right]^{2}\right]^{3/2}}.\label{fx1}\end{equation}
\begin{figure}
\begin{centering}
\resizebox{1\columnwidth}{!}{ \includegraphics{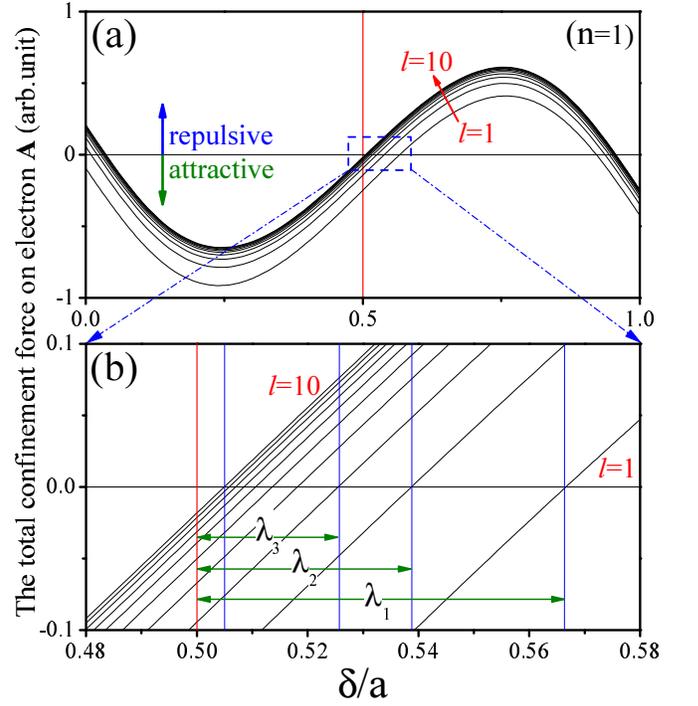}}
\par\end{centering}

\caption{Analytical total confinement force $F_{total}$ on electron $\mathbf{A}$
versus $\delta/a$ for different $l$ with the special condition $Q=+e$
(or $n=1$). Fig. (b) is the enlarged figure of the rectangular area
of Fig. (a).}

\label{fig5}
\end{figure}

Now we can obtain a general formula of the total force $F_{total}$
applied to the electron $\mathbf{A}$ \begin{equation}
F_{total}=f_{B}+F_{x}+F_{x}^{'}.\label{total}\end{equation}

Figure \ref{fig5} shows the relationship between total confinement
force $F_{total}$ and $\delta/a$ for different $l$ with the special
condition $Q=+e$ (or $n=1$). From the Fig. \ref{fig5}(a), it can
be seen clearly that there are two main regions where the right region
indicates the repulsive interaction between electron $\mathbf{A}$
and $\mathbf{B}$, while the left region indicates the attractive
interaction between two electrons. Furthermore, for a given $l$,
as better shown in the enlarged Fig. \ref{fig5}(b) that there exists
a special position $\lambda_{l}$ (the deviation of the electron from
its initial equilibrium position $a/2$) at which the force $F_{total}$
is equal to zero, indicating a complete suppression of the Coulomb
interaction between two electrons. From Fig. \ref{fig4} and Fig.
\ref{fig5}, we can see that as $l$ increases, the corresponding
electron-electron (or stripe-stripe) spacing $\xi$ and the value
($\lambda_{l}$) of zero force position will decrease respectively.

\begin{figure}
\begin{centering}
\resizebox{1\columnwidth}{!}{ \includegraphics{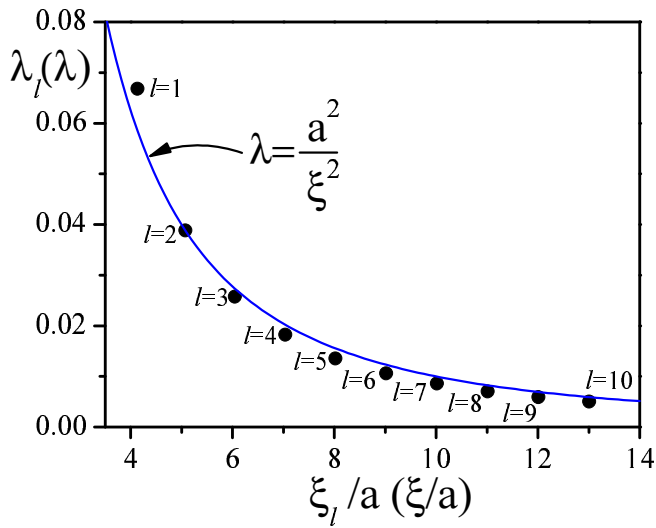}}
\par\end{centering}

\caption{The relationship between $\lambda_{l}$ (the deviation of the electron
from its initial equilibrium position $a/2$) and $\xi_{l}/a$, where
$\xi_{l}$ is the spacing of the electron$\mathbf{A}$ and $\mathbf{B}$.
The relation can be quantitatively described as the following equation
$\lambda=a^{2}/\xi^{2}$, as indicated as the blue line in the figure.}

\label{fig6}
\end{figure}

Moreover, according to the numerical results, we can obtain the $\lambda_{l}$
as a function $\xi_{l}/a$ (where $\xi_{l}$ is the spacing of the
electron $\mathbf{A}$ and $\mathbf{B}$) for electron $\mathbf{A}$
with $l$ from 1 to 10, as shown in Fig. \ref{fig6}. The relationship
between them can be quantitatively described as the following equation:
$\lambda=a^{2}/\xi^{2}$, illustrated as the blue line in Fig. \ref{fig6}.

In our theoretical framework, for a stable superconducting vortex
lattice, the superconducting electrons should develop their self-organized
at the corresponding equilibrium positions ($\lambda_{l}=0$) due
to competition between the stripes. This result implies that the formation
of an ordered superconducting vortex lattice is always accompanied
by the existence of electromagnetic energy ($E_{c}$) inside the superconducting
phase because of the interaction among the vortex lines. This energy
may directly influence the stability of the superconducting vortex
lattice and the corresponding superconducting transition temperature
of the superconductor. In the following, we will discuss qualitatively
which factors might influence the energy $E_{c}$. We know that the
interaction energy between two electrons spaced $\xi$ apart can be
presented by\begin{equation}
E_{e}\propto\frac{1}{\xi}.\label{energy1}\end{equation}
Suppose the distance between two electrons decreases by $\triangle\xi$,
then the interaction energy $E_{e}$ will increase by

\begin{equation}
\triangle E_{e}\sim\frac{1}{\xi^{2}}\triangle\xi.\label{energy2}\end{equation}

From Eq. (\ref{energy2}), it is not difficult to find that the $\nabla E_{e}$
can be effectively reduced by increasing $\xi$ or by decreasing $\triangle\xi$.
In fact, the physical essence of $E_{c}$ can root to the energy of
$\triangle E_{e}$. Physically, we argue that the parameter $\lambda$
of Fig. \ref{fig6} is closely related to the $\triangle\xi$ of Eq.
(\ref{energy3}), for convenience, we assume that they are linearly
related by $\lambda\sim\triangle\xi$. Then according to the above
relation ($\lambda=1/\xi^{2}$) and Eq. (\ref{energy3}), the energy
$E_{c}$ can be simply expressed as:\begin{equation}
E_{c}\sim\frac{1}{\xi^{2}}\triangle\xi\sim\frac{1}{\xi^{2}}\lambda\sim\frac{1}{\xi^{4}}.\label{energy3}\end{equation}

This equation tells us that in order to obtain a more stable superconducting
vortex lattice with a higher temperature, the stripe-stripe spacing
should be sufficiently increased. A more detailed discussion will
be given in the next section.

\begin{table}[b]
\caption{\label{table1}The relationship between the effective c-axis lattice
constant $c^{*}$ and the highest superconducting transition temperatures
$T_{c}^{max}$ which are achieved in the layered cuprate and iron
pnictide superconductors so far.}

\begin{ruledtabular} \begin{tabular}{crcc}
$Materials$  & $c(\mathring{A})$  & $c^{*}(\mathring{A})$  & $T_{c}^{max}(K)$ \tabularnewline
\hline
$*Ba_{1-x}K_{x}Fe_{2}As_{2}$  & $13.212$ & $6.606^{*}$ & $38$ \tabularnewline
$La_{2-x}Ba_{x}CuO_{4}$  & $13.230$ & $6.615^{*}$ & $40$ \tabularnewline
$*(Sr_{4}V_{2}O_{6})Fe_{2}As_{2}$  & $15.673$ & $7.837^{*}$ & $46$ \tabularnewline
$*SmO_{1-x}F_{x}FeAs$  & $8.447$ & $8.447$ & $55$\tabularnewline
$La_{2}Ca_{1-x}Sr_{x}Cu_{2}O_{6}$  & \multicolumn{1}{r}{$19.420$} & \multicolumn{1}{r}{$9.710^{*}$} & $60$ \tabularnewline
$DyBaSrCu_{3}O_{7}$  & \multicolumn{1}{r}{$11.560$} & \multicolumn{1}{r}{$11.560$} & $90$ \tabularnewline
$Tl_{2}Ba_{2}CuO_{6+\delta}$  & \multicolumn{1}{r}{$23.239$} & \multicolumn{1}{r}{$11.620^{*}$} & $92$ \tabularnewline
$YBa_{2}Cu_{3}O_{7-\delta}$  & \multicolumn{1}{r}{$11.676$} & \multicolumn{1}{r}{$11.676$} & $93$ \tabularnewline
$Sr_{2}CuO_{3+\delta}$  & \multicolumn{1}{r}{$12.507$} & \multicolumn{1}{r}{$12.507$} & $95$ \tabularnewline
$Sr_{2-x}Ba_{x}CuO_{3+\delta}$  & \multicolumn{1}{r}{$12.780$} & \multicolumn{1}{r}{$12.780$} & $98$ \tabularnewline
$TlCaBa_{2}Cu_{2}O_{7}$  & \multicolumn{1}{r}{$12.754$} & \multicolumn{1}{r}{$12.754$} & $103$ \tabularnewline
$Tl_{2}CaBa_{2}Cu_{2}O_{8}$  & \multicolumn{1}{r}{$29.318$} & \multicolumn{1}{r}{$14.659^{*}$} & $119$ \tabularnewline
$(Tl_{0.5}Pb_{0.5})Sr_{2}Ca_{2}Cu_{3}O_{9}$  & \multicolumn{1}{r}{$15.230$} & \multicolumn{1}{r}{$15.230$} & $120$ \tabularnewline
$Hg_{2}Ba_{2}Ca_{2}Cu_{3}O_{8}$  & \multicolumn{1}{r}{$15.850$} & \multicolumn{1}{r}{$15.850$} & $133$ \tabularnewline
$HgBa_{2}Ca_{2}Cu_{3}O_{8+\delta}$  & \multicolumn{1}{r}{$16.100$} & \multicolumn{1}{r}{$16.100$} & $136$ \tabularnewline
$Tl_{2}Ca_{2}Ba_{2}Cu_{2}O_{10}$  & \multicolumn{1}{r}{$35.900$} & \multicolumn{1}{r}{$17.950^{*}$} & $128^{*}$ \tabularnewline
$Hg_{2}Ba_{2}Ca_{3}Cu_{4}O_{10}$  & \multicolumn{1}{r}{$19.008$} & \multicolumn{1}{r}{$19.008$} & $126^{*}$ \tabularnewline
\end{tabular}\end{ruledtabular}
\end{table}

\section{the effective c-axis lattice constant and the superconducting critical
temperature}

\begin{figure}
\begin{centering}
\resizebox{1\columnwidth}{!}{ \includegraphics{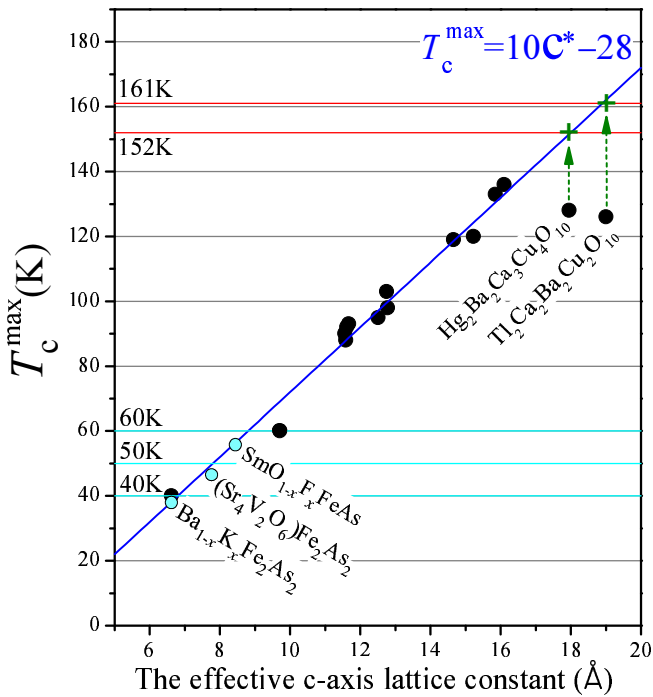}}
\par\end{centering}

\caption{The relationship between the experimental data of the highest superconducting
transition temperatures $T_{c}^{max}$ and the newly defined parameter
of effective c-axis lattice constant $c^{*}$ for the layered cuprate
and iron pnictide superconductors. We argue that the possible maximum
$T_{c}^{max}$ for SmO$_{1-x}$F$_{x}$FeAs of 55 K (1111 series),
(Sr$_{4}$V$_{2}$O$_{6}$)Fe$_{2}$As$_{2}$ of 46 K (21311 series)
and Ba$_{1-x}$K$_{x}$Fe$_{2}$As$_{2}$ of 38 K (122 series) cannot
exceed 60 K, 50 K and 40 K, as indicated by three cyan lines in the
figure.}

\label{fig7}
\end{figure}

Based on the above discussions, we will further explore how to enhance
the superconducting transition temperature of the layered superconductors.
In our scenario, a higher $T_{c}$ means a smaller energy $E_{c}$,
consequently, a larger $\xi$ according to Eq. (\ref{energy3}). It
was pointed out in the previous section, for the doped layered superconducting
materials, there are two ways (the carrier concentration and the c-axis
lattice constant) to adjust the stripe-stripe spacing $\xi$ inside
the superconductors. Our theory suggests that the changing of the
carrier concentration can only modestly increase $T_{c}$, while the
increasing of the effective c-axis lattice constant can lead to a
significant enhancement of $T_{c}$ of the superconducting materials.
This argument has been well confirmed by numerous experimental investigations
in the cuprate and iron pnictide superconductors, as shown in Table
\ref{table1} and Fig. \ref{fig7}.

Table \ref{table1} shows the experimental data of the c-axis lattice
constant $c$ and the highest superconducting transition temperatures
$T_{c}^{max}$ for the layered cuprate and newly discovered iron pnictide
superconductors. Also in Table \ref{table1}, we introduce the effective
c-axis lattice constant $c^{*}$ which can be divided into two different
classes: $c^{*}=c$ and $c^{*}=c/2$ depending on the number of the
superconducting planes inside a unit cell of the superconductors.
The former class indicates that the superconducting layers are separated
by a distance of the c-axis lattice constant, while the superconducting
layer-layer separation along the c-axis is reduced by half for the
later class. Importantly, although the crystal structure and physical
property vary considerably among the superconductors, it is easy to
find that the $T_{c}^{max}$ increases monotonically with the increasing
of the effective c-axis lattice constant.

To obtain the analytical relationship between the $T_{c}^{max}$ and
$c^{*}$, the experimental results of Table \ref{table1} are represented
in Fig. \ref{fig7}. Surprisingly, except for the last two samples
(Tl$_{2}$Ca$_{2}$Ba$_{2}$Cu$_{2}$O$_{10}$ and Hg$_{2}$Ba$_{2}$Ca$_{3}$Cu$_{4}$O$_{10}$),
the figure shows clearly that all of the experimental data almost
fall on a same straight line which can be described with the equation:

\begin{equation}
T_{c}^{max}=10c^{*}-28.\label{temperature}\end{equation}

It should be emphasized that the Eq. (\ref{temperature}) proposed
here has its scope of application since our superconducting mechanism
requires an effective competition among electrons and stripes, too
intensive (for a small $c^{*}$) or too weak (for a large $c^{*}$)
competition is not conducive to the formation of superconducting vortex
lattice. It is estimated from Eq. (\ref{temperature}) that when $c^{*}$
is less than 2.8$\textrm{\AA}$, the corresponding layered materials
can not exhibit the superconductivity. Of course, there must be a
maximum value of $c_{max}^{*}$ which also led to the failure of the
establishment of the superconducting vortex lattice inside the materials.
We note from the figure that, as unexpected, the superconducting transition
temperatures of Tl$_{2}$Ca$_{2}$Ba$_{2}$Cu$_{2}$O$_{10}$ and
Hg$_{2}$Ba$_{2}$Ca$_{3}$Cu$_{4}$O$_{10}$ deviate seriously from
the linear equation of Eq. (\ref{temperature}). We think that there
may be two reasons for these deviations. First, since the carrier
concentrations of these two superconductors are unadjustable, it is
likely that the vortex lattices are not in best consistent with those
of the crystal structures. Second, their effective c-axis lattice
constants have been very close to the maximum value of $c_{max}^{*}$.
Personally, we believe that the maximum $T_{c}^{max}$ of the cuprate
superconductors may be raised to the theoretical value (about 161
K, as marked in Fig. \ref{fig7}) without applying of the external
pressure. Then, can we predict the limits of the $T_{c}^{max}$ according
to our theory? From the viewpoint of lattice stability and competition
among superconducting electrons, a maximum value of $c_{max}^{*}$
values range from 19$\textrm{\AA}$(the $c^{*}$ of Hg$_{2}$Ba$_{2}$Ca$_{3}$Cu$_{4}$O$_{10}$
) to 19+3$\textrm{\AA}$ (where 3$\textrm{\AA}$ is about the thickness
of two atomic layers). Hence by Eq. (\ref{temperature}), we conclude
that the limits of the $T_{c}^{max}$($\sim192$K) is very difficult
to break through 200 K in three-dimensional bulk superconductors.

As is well known, soon after the discovery of the iron-based superconductors,
some physicists have believed room temperature superconductivity may
be possible in these new compounds. We think that these predictions 
are apparently lack of solid scientific basis. As can be found from
Table \ref{table1} and Fig. \ref{fig7}, the experimental data of
three iron-based superconductors: SmO$_{1-x}$F$_{x}$FeAs of 55 K
(1111 series),\cite{ren} Ba$_{1-x}$K$_{x}$Fe$_{2}$As$_{2}$ of
38 K (122 series)\cite{marianne} and (Sr$_{4}$V$_{2}$O$_{6}$)Fe$_{2}$As$_{2}$
of 46 K (21311 series)\cite{zhu} According to the theory suggested
in this paper, we argue that the possible maximum $T_{c}^{max}$ for
1111, 122 and 21311 series cannot exceed 60 K, 40 K and 50 K, as indicated
in Fig. \ref{fig7} (three cyan lines).

Furthermore, we will show that our theory can provide a qualitative
explanation of why the superconductivity does not occur in noble metals
(for example, Ag, Au, and Cu) and a also new insights into the problem
of the superconductivity in metallic hydrogen. It is a common knowledge
that gold, silver and copper are the best conductors of electrical
current. However, a question has long plagued the physics community:
why an ideal conductor that is not a superconductor? With the help
of the Fig. \ref{fig2} and Eq. (\ref{temperature}), we try to unravel
this mystery in a very simple way. Under normal circumstances, good
conductors always have an extremely high level of carrier concentration.
One can assume that these metals have the superconducting properties, 
in our view, 
it is necessary that the superconducting electrons should be arranged
as the specific vortex lattices of Fig. \ref{fig2} with the stripe-stripe
spacings $\xi\sim2\textrm{\AA}$ ($\ll2.8\textrm{\AA}$). In this
case, the crowded vortex lattices are unstable owing to the strong
electromagnetic interactions between vortex lines, as a result, the
charge carriers are more likely to be formed in a random and lower-energy
stable non-superconducting phase. These discussions can also be applied
to the overdoped high-$T_{c}$ superconductors where the stripe-stripe spacing inside 
the superconducting CuO planes is far less than $2.8\textrm{\AA}$. 

Physicists predicted
that hydrogen in solid-state can be a high-temperature superconductor,
and maybe even a room-temperature superconductor. Indeed, this prediction
sounds very interesting and attractive, but we would like to point
out that it cannot be realized as desired. From the perspective of
atomic structure, there are the same number of carriers (free electrons)
in metallic hydrogen, lithium (BCC, $a=b=c=3.51\textrm{\AA}$), sodium
(BCC, $a=b=c=4.29\textrm{\AA}$) and potassium (BCC, $a=b=c=5.33\textrm{\AA}$)
with the same number of atoms. The theoretical predicted lattice constants
of metallic hydrogen (BCC, $a=b=c=2.89\textrm{\AA}$) are smaller
than those of the other three materials, indicating that metallic
hydrogen has the highest carrier concentration among them. As lithium,
sodium and potassium do not exhibit superconductivity, now we can
boldly predict that the metallic hydrogen with the highest carrier
concentration cannot be a superconductor at any low temperature, not
to mention the room temperature superconductivity.

\begin{figure}
\begin{centering}
\resizebox{1\columnwidth}{!}{ \includegraphics{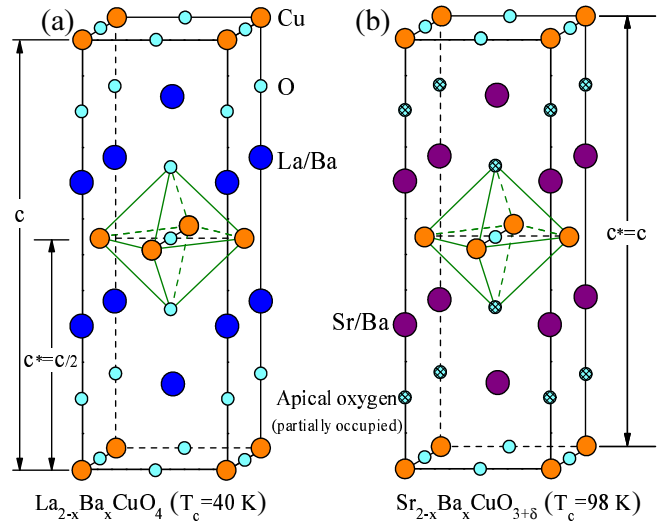}}
\par\end{centering}

\caption{Two analogous superconductors (a) La$_{2-x}$Ba$_{x}$CuO$_{4}$ of
40 K, and (b) Sr$_{2-x}$Ba$_{x}$CuO$_{3+\delta}$ of 98 K. In our
theoretical framework of the effective c-axis lattice constant, they
have completely different relation between the c-axis lattice constant
$c$ and the effective c-axis lattice constant $c^{*}$: $c^{*}=c/2$
for La$_{2-x}$Ba$_{x}$CuO$_{4}$ with two superconducting planes inside one unit cell,
while $c^{*}=c$ for Sr$_{2-x}$Ba$_{x}$CuO$_{3+\delta}$ with only one superconducting plane inside one unit cell.
We think it is the difference of the effective c-axis lattice constant
that leads to the very difference between two samples in superconducting
transition temperature.}

\label{fig8}
\end{figure}

Finally, it is worth to note that the suggested real-space effective
c-axis lattice constant theory of superconductivity has been excellently
confirmed by recent measurements.\cite{wbgao} Fig. \ref{fig8}(a) shows a unit-cell
of cuprate La$_{2-x}$Ba$_{x}$CuO$_{4}$of 40 K with the c-axis lattice
constant $c=13.23\textrm{\AA}$. It is easy to find from the figure
that superconducting layer (CuO plane) spacing (the effective c-axis
lattice constant) of the superconductor is $c^{*}=c/2=6.615\textrm{\AA}$.
If there are experimental methods that can cause one of the CuO plane
to lose the ability of conducting superconducting current, then the
effective c-axis lattice constant can be promoted to $c^{*}=c=13.23\textrm{\AA}$,
consequently, the corresponding maximum $T_{c}^{max}$ may be dramatically
enhanced to 104 K calculated according to Eq. (\ref{temperature})
. This prediction has just been experimentally verified by Gao et
al.\cite{wbgao} in cuprate Sr$_{2-x}$Ba$_{x}$CuO$_{3+\delta}$
which look very similar to that of La$_{2-x}$Ba$_{x}$CuO$_{4}$
but with partially occupied apical oxygen sites, as shown in Fig.
\ref{fig8}(b). Why can two almost the same superconducting materials
have completely different superconducting transition temperatures?
Although several factors have been considered in what enhance the
$T_{c}$,\cite{wbgao} it is intuitively obvious that the actual reason
has not been elucidated yet. In our theoretical framework of the effective
c-axis lattice constant, it is most likely that the introducing of
partially occupied apical oxygen sites could cause one of the CuO
layers no longer to function as the superconducting plane. Therefore,
the two analogous superconductors of Fig. \ref{fig8} have completely
different relation between $c$ and $c^{*}$: $c^{*}=c/2$ for La$_{2-x}$Ba$_{x}$CuO$_{4}$
of Fig. \ref{fig8}(a), while $c^{*}=c$ for Sr$_{2-x}$Ba$_{x}$CuO$_{3+\delta}$
of Fig. \ref{fig8}(b). From the experimental results of Sr$_{2-x}$Ba$_{x}$CuO$_{3+\delta}$
($c^{*}=c=12.78\textrm{\AA}$)\cite{wbgao} and Eq. (\ref{temperature}),
one can get immediately $T_{c}^{max}\approx99.8$ K which is in good
agreement with the experimental result ($T_{c}^{max}\approx98$ K).\cite{wbgao}

\section{Concluding remarks}

We have proposed the real-space effective c-axis lattice constant
theory of superconductivity based on the self-organized picture of
the superconducting electrons and studied the stability of the superconducting
vortex lattices in layered superconductors. It has been shown clearly
that the effective lattice constant play a significant role in promoting
the superconducting transition temperature in layered superconductors,
such as the copper-based and newly discovered iron-based compounds.
According to a large number of experimental data, we have obtained
an important equation: $T_{c}^{max}=10c^{*}-28$ which can be used
to estimate the highest possible $T_{c}^{\max}$ for a given layered
superconductor, where $c^{*}$ is the effective c-axis lattice constant.
This result suggests that the maximum possible $T_{c}^{max}$ of the
iron-based superconductors cannot exceed 60 K, 50 K and 40 K for the
1111, 21311 and 122 series, respectively. It should be noted that
this prediction has stood the test of time for about two years by
many experiments. Furthermore, we have tried to explain why two very
similar cuprate superconductors (La$_{2-x}$Ba$_{x}$CuO$_{4}$ of
40 K and Sr$_{2-x}$Ba$_{x}$CuO$_{3+\delta}$ of 98 K) would have
such a different superconducting transition temperatures, the latter
is about 2.5 times of the former. The physical reason why the superconductivity does not occur
in gold, silver and copper has also been provided based on the suggested
new mechanism. Finally, we have argued that the metallic
hydrogen cannot exhibit superconductivity at any low temperature.
We think that the real-space effective c-axis lattice
constant theory of superconductivity may finally shed light on the
mysteries of superconductivity. With these results, the scientists
will be able to proceed to the materials design for the new superconductors
that may have a higher superconducting transition temperature.

\end{document}